\documentclass[12pt]{iopart}

\newcommand{\claire}[1]{\color{black}#1\color{black}}
\newcommand{\co}[1]{\color{black}#1\color{black}}
\newcommand{\matthew}[1]{\color{black}#1\color{black}}

\newcommand{\mattfinal}[1]{\color{black}#1\color{black}}

\usepackage{ulem}
\usepackage{bm}
\usepackage{amssymb}  
\usepackage{amsmath}  
\usepackage{leftidx}
\usepackage{graphicx}
\usepackage{enumitem}
\usepackage{hyperref}
\usepackage[capitalise]{cleveref}
\usepackage[dvipsnames]{xcolor}
\newcommand{\eq}[1]{\begin{align}#1\end{align}}
\newcommand{\cmmt}[1]{\color{black}#1\color{black}}

\newcommand{\matt}[1]{\color{black}#1\color{black}}

\begin{document}


\title{Sensitivity Analysis for 
\matt{Optimizing } Electrical Impedance Tomography Protocols}


\author{Claire Onsager*, Chulin Wang*, Charles Costakis*, Can Aygen*, Lauren Lang*, Suzan van der Lee† and Matthew A. Grayson*
}

\address{*Department of Electrical and Computer Engineering, Northwestern University, Evanston, IL\\
† Department of Earth and Planetary Sciences, Northwestern University, Evanston IL
}
\vspace{10pt}
\begin{indented}
\item[]November 2021
\end{indented}

\begin{abstract}
Electrical impedance tomography (EIT) is a
noninvasive imaging method whereby electrical measurements on the boundary of a conductive medium \matthew{(the data) are taken according to a prescribed protocol 
set and inverted } 
to map the internal conductivity (the model).  \matt{This paper introduces a sensitivity analysis method } \matthew{and corresponding inversion and protocol optimization that generalizes the criteria for tomographic inversion to } \matt{ 
minimize the model-space dimensionality and maximize data importance. }
\matt{Sensitivity vectors, defined as rows of the Jacobian matrix in the linearized forward problem, } \matthew{are used to }
\matt{map } \matthew{targeted } \matt{conductivity features from model-space to data-space, } \matthew{and a volumetric outer-product of these vectors in model-space called the sensitivity parallelotope volume }
provides a figure-of-merit \matthew{for data protocol optimization. }
\matt{Orthonormal basis functions that accurately constrain the model-space to features of interest can be defined from  \textit{a priori} information.
By increasing the contact number } \matthew{to } \cmmt{
expand the }\mattfinal{number of possible measurements 
$D_\textrm{max}$, } 
\matt{and by reducing the model-space to } \cmmt{a minimal number $M_0$ of } \matt{basis functions that describe only the features of interest, the } 
\matthew{$
M_0 \ll D_\mathrm{max} $ }
\matthew{
sensitivity vectors of greatest length }\mattfinal{and maximal orthogonality }\matt{ 
that 
span this model-space can be identified. }
\matt{The reduction in model-space dimensionality }\matthew{accelerates the inversion by several orders of magnitude, } \matt{and the } \matthew{enhanced sensitivity can } 
\matt{tolerate noise levels up to 1,000 times larger than standard } \cmmt{protocols.}

\end{abstract}

%
%
%
%
%

\section{Introduction}

Electrical impedance tomography (EIT) of conducting volumes holds great promise as an alternative to X-ray computed tomography (CT) and MRI with applications in medicine, engineering, and geoscience, but is not commonly used because of its diffuse nature. 
The disadvantage of \matthew{EIT's } 
diffuse nature can be compensated 
\matthew{by the advantages of being } radiation-free, easily portable, wearable, and low-cost.
\matthew{But t}wo longstanding challenges have prevented EIT from realizing rapid refresh rates and being robust against \claire{experimental } noise: \matt{First, the computational time of existing algorithms increases geometrically with the number of model-space parameters, and }\mattfinal{the ill-posed nature of the inverse problem under }\matt{standard algorithms } \matthew{requires } 
\matt{a dense mesh of up to 10,000 such model parameters.  Thus fast EIT refresh times become possible only at the cost of reducing resolution. } 
The second challenge is that 
present-day EIT \matt{relies on measurement protocols that include small-signal measurements that are vulnerable to }\claire{experimental }\matt{noise.  Error in these small-signal measurements will unavoidably impair the fidelity of the inverse problem, thereby reducing the resolution.  In this work, we introduce a sensitivity analysis which can accelerate the computational time } \matthew{by orders of magnitude } \matt{by 
reducing the model-space } \matthew{to only the 
resolution of interest, } \matt{and improve the robustness to noise } \matthew{by orders of magnitude } \matt{by eliminating }\mattfinal{low sensitivity }
\matt{measurements } \matthew{from the EIT protocol.}

The remainder of this paper is structured as follows. 
\cref{sec:review} 
will review EIT concepts and introduce the notation adopted here. 
\cref{sec:model_space} \matt{introduces the reduced 
model-space, 
chosen to have only as many elements as are needed to represent } \matthew{the desired resolution or } \matt{the features of interest. } \matthew{\cref{sec:contact_allocation} explains how many contacts are needed to benefit from the sensitivity analysis method. }\matt{ 
\cref{sec:measurement_space} introduces the measurement-space whereby 
} \matthew{a deliberately large   } \matt{number of contacts is 
chosen to exceed the minimum necessary to map the model-space. And \cref{sec:data_space} introduces the data-space 
} \matthew{that is chosen from this vast measurement-space } \matt{
to maximize sensitivity and 
linear independence. Finally, \cref{sec:inverse_sol} } \co{introduces two iterative Newton's method approaches to solve the inverse problem }
\matthew{under this sensitivity paradigm. } \matt{ \cref{sec:results} of this work will show the results of simulations that apply these methods to different applications. An outline overview of the sensitivity enhanced measurement } \matthew{analysis and } \matt{ protocol is } \matthew{provided } \matt{
below: }

\begin{enumerate}

   \item  {\bf Model-Space} (Sec.~\ref{sec:model_space})
   \begin{enumerate}
     \item \underline{Reference model} based on prior knowledge
     \item \underline{Orthonormal basis} based on features or resolution desired 
   \end{enumerate}
   \item{\bf Contact Allocation} (Sec.~\ref{sec:contact_allocation})
   \item {\bf Measurement-Space} (Sec.~\ref{sec:measurement_space})
   \item {\bf Data-Space} (Sec.~\ref{sec:data_space})
   \begin{enumerate}
     \item \underline{Pruning criteria} for measurement-space
     \item \underline{Optimal measurement protocol} selection:    \begin{enumerate}
         \item Maximize sensitivity parallelotope volume
        \item Other protocol optimization criteria
      \end{enumerate}
    \end{enumerate}
    \item {\bf \co{Inverse Solution for Sensitivity Analysis }}
    (Sec.~\ref{sec:inverse_sol})
   \begin{enumerate}
     \item \underline{\mattfinal{(Truncated) }\co{Singular Value Decomposition}} Inverse Solving
     \item \underline{\co{Regularized}} Inverse Solving
    \end{enumerate}
 \end{enumerate}

\section{Standard Electrical Impedance Tomography Method and Protocols}
\label{sec:review}

\matt{The essential components of electrical impedance tomography (EIT) will be reviewed here to establish the notation used in the remainder of the paper.  The } \matthew{generalized forward } \matt{
problem 
\begin{equation}
    \label{eq:forward_problem}
    \bm{d} = \bm{F}(\bm{m})
\end{equation} 
relates } 
\matthew{a vector $\bm{m}$ in } 
model-space, \matt{whose basis vectors describe } 
the conductivity distribution, to 
\matthew{a vector $\bm{d}$ in } 
data-space, \matt{whose basis vectors describe } 
the measurements taken. \matt{  Typically, } \matthew{this } 
\matt{forward problem 
can be directly solved } using Maxwell's equations and various boundary conditions, 
and details 
can be found in the book ''Electrical Impedance Tomography: Methods, History and Applications'' by Holder \cite{bible}. 

EIT is \matt{classified as  
an inverse problem, whereby the data vector $\bm{d}$ is given, and the model vector $\bm{m}$ needs to be solved } \claire{for}. \matt{In the presence of } \claire{experimental }\matt{data noise $\bm{n}$, }
the observed data 
$\bm{d^{\textrm{obs}}}$ 
can be written as:
\eq{\bm{d^{\textrm{obs}}}&=\bm{F}(\bm{m})+\bm{n}.}
\matt{The forward operation $\bm{F}(\bm{m})$ can be Taylor expanded to linear order, }
\eq{\label{eq:matrix_element}
J_{\alpha,\beta}& =\frac{\partial F_\alpha}{\partial m_\beta},}
\matt{where $\alpha \in \{ 1,\cdots D_0\}$ indexes the data values \cmmt{from 1 to the rank of the data vector $D_0$ 
\matthew{(the dimensionality of the data-space), } }
and $\beta \in \{ 1,\cdots M_0\}$ indexes the model elements \cmmt{from 1 to the } \matthew{rank of the model vector $M_0$ (the } \cmmt{dimensionality of the model-space). } 
}
 The forward problem can then be represented by \claire{the } Jacobian matrix 
\eq{
\bm{d^{\textrm{obs}}}&  =\bm{J m}+\bm{n},
\label{eq:jacobian_forward}}
 \matthew{where } each element of the Jacobian \matt{quantifies } the change in a data-space element with respect to a change in a model-space component. \matt{If the }\claire{magnitude of a } \matt{Jacobian matrix element is large, then small changes in the corresponding model-space component will result in large swings in the observed data, }\matthew{implying high {\em sensitivity} of that particular measurement to that feature in model-space.  It is precisely this concept of sensitivity which will inspire the figure-of-merit for optimizing measurement protocols in Section\,\ref{sec:sensitivity_analysis}. }

\subsection{The Model-Space }
\label{sec:intro_model_space}
In most EIT \matt{treatments, } 
the model-space \matt{is represented with } 
a finite element mesh. 
\matt{The number of independent model elements will be indexed $b \in \{1\cdots M_\mathrm{max}\}$, where $M_\mathrm{max}$ is the number of independent mesh elements, } 
\mattfinal{and } \matt{ $M_\mathrm{max}$ must be
sufficiently large for the discrete mesh to approximate a } \cmmt{continuum } \matt{solution to Maxwell's equations. But } in order to solve the inverse problem, the conductivity must be determined for each \matt{ of these $M_\mathrm{max}$ } mesh element\matt{s, and } 
the number of unique \matt{data values $D$}\matthew{$_0$ }\matt{ measured at the }\claire{sample } \matt{boundary } 
is much less than the number of mesh points \matt{$D$}\matthew{$_0$}\matt{$ \ll M_\mathrm{max}$}. 
The resulting inverse problem is greatly under-determined and has no unique solution, produc\matt{ing instead } many equally likely conductivity maps for a given set of boundary \matt{data values.  
Thus, } inverse solving methods 
\matt{typically } implement regularization \matthew{techniques } \matt{to artificially select one solution from among the others, usually prioritizing smoothness with a weighting term in the regularization.} 

As an alternative to \matt{representing } 
model-space 
\matt{with a large number $M_\mathrm{max}$ of mesh }\claire{pieces, }
\matt{one can instead represent it with } a reduced \matt{number } $M_0$ 
of orthonormal basis functions \cmmt{that are } \matt{defined over the mesh } 
\matthew{whereby } 
\matt{$M_0 \ll M_\mathrm{max}$. }  
Conductivity maps \matt{are }\matthew{then }\matt{ represented as } 
linear superpositions of \matt{these } individual basis functions, and prior knowledge of the problem is used to \matt{select appropriate basis functions. } \cmmt{These basis functions thereby } reduc\cmmt{e }
model-space dimensionality, and 
\matt{reduc}\cmmt{e } \matt{or eliminat}\cmmt{e the } under-determined nature of the problem.  

\matt{A } reduced dimensionality model-space \matt{has the advantage of } greatly reduc\matt{ed } 
computational time. \matt{For example, any EIT inversion requires } 
Jacobian solvers 
\matt{and their computation time is empirically observed to } scale 
approximately \matt{as } \claire{$M_0^{3/2}$ } with the number of model-space elements \claire{$M_0$ } \cite{Mscale}. 
One way to 
reduce the size of the model-space is the approach taken by Vauhkonen {\it et al.},  who created orthonormal basis functions for thorax EIT imaging using anatomical information and a proper orthogonal decomposition (POD) of previously collected data \cite{basis}. A similar approach was taken \matthew{by } 
Lipponen {\it et al.} \matthew{who } used POD to produce orthonormal basis functions for the location of circular inhomogeneities within a sample of otherwise homogeneous conductivity \cite{POD}. 
The resulting basis functions show strong resemblance to orthonormal polynomial basis functions \matt{defined over a circular domain } called Zernike polynomials \cite{momentproblem}. \matt{Together, these results  
suggest that for a circular sample in the absence of data for } proper orthogonal decomposition, 
the Zernike polynomial functions \matt{provide an excellent empirical basis. } 
When selecting a subset of these polynomials to use as \claire{a } basis, \matt{Vauhkonen {\it et al.} recommend using } 
a number $M_0$ less than or equal to the amount of information being measured \matt{$M_0\le D$}\matthew{$_0$ } to reduce the ill-posed nature of the problem \cite{basis}. This recommendation is supported by Tang {\it et al.}, whose work indicates that one should also consider any knowledge of relative feature location when selecting basis functions \cite{degreesOfFreedom}.

\subsection{The Data-Space}
\label{sec:intro_data_space}
\mattfinal{Each four-point resistance measurement corresponds to a single data element in the }  \matthew{
data vector to be inverted in the tomographic problem. } 
\matthew{From the $C$ available contact electrodes, each resistance measurement will be made with two 
current and two voltage contact pairs, constituting a four-point measurement, often called a ``tetrapolar" } \matthew{resistance measurement } \co{in the biomedical device community \cite{tetrapolar}. }
\matt{
Adopting } 
the van der Pauw notation \cite{vdp}, the current and voltage contacts of each four-point measurement \matt{are designated } $[I_+, I_-, V_-, V_+]$. \matt{Though some EIT efforts 
consider 
two-point measurements } \cite{Zhang} \cite{Ma},  
\matt{the analysis here will be restricted to } \matthew{the } \matt{four-point } \matthew{configuration } 
\matt{since }
\matthew{it } \matt{
tolerates finite contact impedances common to most experimental applications, such as in biomedicine } \cite{BIOEIT}.
\matthew{When a particular {\it set} of $D_0$  data measurements is designated for tomographic inversion this sequence }
\matt{will } 
\matthew{be called 
a measurement {\em protocol}. }

\matt{There is a large number of possible four-point measurement } \matthew{protocols } 
\matt{for a given number of contacts $C$.  There are $\binom{C}{4}$ different ways of selecting a particular combination of 4 contacts $i,j,k,l$, and there are three 
\claire{unique } four-point resistance values that can be measured with these four contacts, for example, $[i,j,k,l], [j,k,l,i],$ and $[i,k,j,l]$.
All other permutations of these same 4 contacts will yield either the same resistance value due to the Onsager reciprocity relation } \matt{(whereby resistance values are 
equal when current and voltage pairs are swapped) \cite{onsager} or the negative thereof (due to polarity switching.) } \matt{  It is worth noting that, of the 
three measurement } \matthew{values } \matt{above, only two are linearly independent, and the third one is a linear combination of the other two.  Thus the total number of data  measurements that yield different, but not necessarily independent, resistance values is}: 
\eq{
    D_\mathrm{max}=3\binom{C}{4} = \frac{3 C!}{4!(C-4)!}.
    \label{eqn:Dmax}
}
\matthew{This 
number 
$D_\mathrm{max}$ includes many 
resistance values that are not linearly independent -- such as when two measurements share three contacts, $[i,j,k,l]$ and $[i,j,l,m],$ and the resistance measurement $[i,j,k,m]$ returns a value that is 
merely the sum of the first two. } \matthew{To make }\matt{
every possible measurement would not only be impractical with the number of measurements } \matthew{ $D_\mathrm{max}$ } \matt{diverging as $O(C^4)$, it }
\matthew{would } \matt{also } \matthew{be } \matt{ entirely unnecessary given the amount of redundant information } \matthew{ described above. } \matt{  Thus, a }\matthew{typical } \matt{measurement protocol will always concern a }\matthew{smaller data } \matt{ subset $D_0$ of these  measurements such that $D_0 \ll D_\mathrm{max}$. }

\matthew{The number $D_\mathrm I$ of possible {\em independent} measurements for $C$ contacts } \claire{\cite{sheffield} } \co{is important in the classification of a protocol as redundant, efficient, or compliant: }
\matthew{ 
\begin{equation}
    D_\mathrm{I}=\frac{C(C-3)}{2}.
    \label{data_independent}
\end{equation} 
If the number of measurements in a protocol $D_0$ is less than the number of independent measurements $D_0 < D_\mathrm{I}$ then we will describe the protocol as ``compliant," meaning that there is independent information available that is not being measured.  If, on the other hand $D_0 > D_I$, then we shall call such a protocol ```redundant," since more measurements are taken than necessary to acquire all the independent data.  Finally, the case where all possible independent data measurements are represented in the protocol $D_0 = D_\mathrm{I}$, shall be called ``efficient"}.
\co{The above definitions are summarized in Table \ref{tbl:classtable}. } 

\begin{table}
\mattfinal{
\begin{eqnarray}
 &\mathrm{\underline{Protocol~Classification} }  \nonumber\\
 \mathrm{Redundant: } & ~ D_0 > D_\mathrm{I},  ~~~ D_0 = r D_\mathrm{I} ~~~(r, \mathrm{redundancy~factor}) \nonumber\\ 
 \mathrm{Efficient: } & ~ D_0 = D_\mathrm{I},  ~  \nonumber\\  
 \mathrm{Compliant: } & ~ D_0 < D_\mathrm{I},  ~~~ c D_0 = D_\mathrm{I}~~~ (c, \mathrm{compliance~factor}\nonumber) 
\end{eqnarray}
\caption{Classification of protocols in terms of the amount of data collected.  $D_0$ is the number of measurement data collected; $D_\mathrm{I}$ is the number of independent data possible.}
}
\label{tbl:classtable}
\end{table}

\matthew{Redundant protocols can be helpful in noise reduction, since in the presence of experimental noise, a redundancy factor $r$ defined as $D_0 = r D_\mathrm{I}$ will reduce the overall measurement noise by a factor of $\sqrt{r}$. However, both redundant and efficient protocols include by necessity many } \claire{low sensitivity }\matthew{measurements which, in the presence of experimental noise, can be the primary source of errors in the tomographic inversion. Therefore, a key strategy of the present work is to recognize that compliant protocols have the advantage of permitting a subselection of only the } \claire{highest signal } \matthew{measurements for the protocol. Thus, compliant protocols such as those introduced in Section\,\ref{sec:sensitivity_analysis} have the potential to provide the greatest robustness to experimental noise, as will be demonstrated in the simulated tomographic inversions in Section\,\ref{sec:results}. }

\matthew{In choosing independent measurements to comprise a protocol, it is useful to identify 
which measurements } \claire{are relatively insensitive to changes in model-space, and which measurements use similarly positioned contacts so as to }\mattfinal{be sensitive }
\claire{to largely the same features. } 
\matthew{
}\claire{Low sensitivity } \matthew{measurements can be identified by inspection as measurements where either the current or voltage terminal pairs (or both) neighbor each other}, \claire{often referred to as ``adjacent" measurements \cite{adjacent}}.  \matthew{If the current pairs are neighboring, the voltage is dropped only over a small region of the sample between those two contacts where no voltage contacts are available to measure, and if the voltage terminals are neighboring, small voltage amplitudes are measured.  One would naively think that the best protocols with the strongest signals would then require that one avoid neighboring current or voltage pairs.  However, the second problem of similar measurement pairs comes into play.
Consider a }\claire{low sensitivity } \matthew{measurement $[i,j,l,m]$ where, for example, voltage contacts $l$ and $m$ are neighboring each other}. \claire{This low sensitivity measurement can be added to a high sensitivity } \matthew{measurement $[i,j,k,l]$ to yield the } \claire{high sensitivity } \matthew{measurement $[i,j,k,m]$.  Although both $[i,j,k,l]$ and $[i,j,k,m]$ }\claire{have high sensitivity and are }\matthew{linearly independent from each other, their similarity aside from a } \claire{small sensitivity }\matthew{measurement means that they effectively provide no stronger a }\claire{measurement } \matthew{than if the }\claire{low sensitivity measurement }\matthew{were, itself, in the protocol instead of either of these. 

We will now examine some protocols and qualify them according to their compliant, efficient, or redundant character, as well as whether they are prone to }
\claire{low sensitivity } \matthew{measurements 
as described above. }
\matt{In the published literature, } \matthew{most } 
\matt{EIT studies use the Sheffield protocol to take four-point resistance measurements \cite{sheffield}. 
Under } this 
\matt{protocol, the }\matthew{number } \matthew{of measurements is }\matt{ $D_\mathrm{S} = C(C-3) = 2 D_\mathrm{I},$ }
\matthew{which are taken }\matt{``pairwise } adjacent" whereby 
two current contacts neighbor one another and 
two voltage contacts neighbor one another. 
Starting with a given current pair, the \matt{nearest } voltage \matt{pair is measured, and then the voltage pair } 
rotated around the sample \matt{for each subsequent measurement until } 
all \matt{remaining contact pairs have been measured. Then, the current pair is rotated by one step and the process repeated } \cite{sheffield}. 
For $C$ contacts \matt{indexed counter-clockwise around the sample, the full Sheffield protocol is tabulated below for the first current pair in the first column, the second current pair in the second column, } {\it etc.}:
\begin{equation}
\begin{matrix}
    [1,2,3,4]& [2,3,4,5] & [3,4,5,6] & \cdots & [C,1,2,3]\\
    [1,2,4,5]& [2,3,5,6] &  \vdots   &        & \vdots\\
    [1,2,5,6]&   \vdots  &  \vdots   &        & \vdots\\
      \vdots &   \vdots  &  \vdots   &        & \vdots\\
  [1,2,C-1,C]& [2,3,C,1] & [3,4,1,2] & \cdots & [C,1,C-2,C-1]
\end{matrix}
\end{equation}
The \matthew{$C$-columns times $(C-3)$-rows in the table above confirm the number of measurements $D_\mathrm{S}$ for the } Sheffield protocol.  \matthew{The protocol redundancy factor of $r = 2$ arises } 
\matt{due to the Onsager reciprocity relation, }
\matthew{whereby the top row is redundant with the bottom row, the second row with the second-to-last row, {\it etc. } } 
\matthew{When this redudancy is removed, the resulting efficient } protocol \matthew{with $D_\mathrm{I}$ measurements } will be 
referred to 
as the {\em reduced} Sheffield protocol.  
\matt{The empirically derived Sheffield protocol has the advantage of }\matthew{being easy to implement while } \matthew{treating all contacts symmetrically and has withstood the test of time in that no significantly superior } \matthew{redundant (or efficient) protocol }
\matt{in 2D has been determined, especially when no prior knowledge exists for the system under study. }

Other \matt{heuristic } protocols have been proposed, \matthew{for example, } throughout a series of papers by Adler to satisfy external measurement constraints \cite{GREIT}, \cite{adjacent}. One of these suggests, for example, that adjacent measurements are not always ideal, especially for medical applications where low current levels are medically safer. \co{Adler's publications developed 9 problem-specific figures-of-merit for trial and error comparisons of measurement protocols \cmmt{\cite{Graham} \cite{GREIT} \cite{adjacent} \cite{Mamatjan} \cite{3DGREIT} \cite{Wagenaar} \cite{Thurk}}.  }
However, \matthew{
as mentioned previously, all such efficient or redundant protocols suffer from the same drawback:  } 
\matt{they include by necessity 
either }\claire{low sensitivity measurements } \matthew{or, equivalently, measurement pairs which differ by a }\claire{low sensitivity measurement, } 
\matt{ since all of these protocols seek to exhaust as many independent measurements as possible for a given number of contacts. As a result, such protocols can be susceptible to } \matt{ low noise levels. } 
\matthew{As a final note, }
\matt{the measurement configurations proposed for these heuristic protocols are chosen in a somewhat {\it ad hoc} manner through trial and error, rather than according to a mathematically rigorous figure-of-merit. 
Section\,\ref{sec:sensitivity_analysis} 
proposes sensitivity 
as an appropriate figure-of-merit to identify the optimal }\matthew{compliant }\matt{protocol 
from 
an increased space of possible measurements with increased contact number. }  
\matthew{The resulting } \matt{sensitivity protocol is }\matthew{expected to be }\matt{more robust to noise, }\matthew{while being mathematically justifiable through a figure-of-merit. }

\color{black}
\matt{Other recent EIT efforts have tried to increase EIT performance }
\co{through the selection of electrode placement and measurement protocols. For example, }
Karimi describes how variations in electrode placement can maximize “information gain” \co{by using a double-loop Monte-Carlo approximation \cite{Karimi}, and }
Smyl selected low-noise electrode positions for the 2D problem through deep learning \matt{with prior knowledge of the system under study } \co{\cite{Smyl}. }
\co{The methods the sensitivity protocol uses for the measurement protocol optimization problem can also be applied to improve this electrode position optimization problem. }Recent work by Ma uses a reduced selection of model-space features and measurement protocols via Bayesian methods for hand gesture recognition, \co{using their methods to distinguish between 11 different gestures \cite{Ma}. } \co{Section \ref{sec:results} demonstrates how our sensitivity protocols show improved distinguishabilty in the presence of experimental noise. }
\subsection{Linear and Nonlinear Solutions}
\label{sec:intro_inverse_sol}
The methods to solve inverse problems fall under one of two categories, linear or nonlinear. 
In the linear problem introduced in \cref{eq:jacobian_forward}, 
the forward problem 
is treated as \matt{simple } matrix multiplication. 
The corresponding inverse problem can be written as:
\eq{\bm{m}&  =\bm{J}^{-g} (\bm d+\bm n),
\label{eqn:inverse}}
where $\bm{J}^{-g}$ is a generalized inverse of the Jacobian. In the case where 
model and data-spaces \matt{have the same size } 
\cmmt{$M_0=D_0$}, 
the Jacobian is square and the inverse of the Jacobian matrix $\bm{J}^{-1}$ can be \cmmt{substituted ($\bm{J}^{-g}=\bm{J}^{-1})$}.

\matt{However, the general } EIT inverse problem is 
\matt{inherently } nonlinear. One approach to solve this nonlinear problem is 
\matt{iteratively, } such as \matt{with } Newton's method, \matt{which }
solves the nonlinear problem by solving a linear 
\matt{approximation } at each iteration. For example, \matt{from } the Jacobian \matt{of }\matthew{a } \matt{
previous iteration } the data residuals \matt{of the present iteration }\matthew{can be }\matt{
calculated } at each step. Each iteration is treated as an update to the 
\matthew{one prior, } starting with the problem reference as the initial ``guess''. The iterative process stops once this update\matt{d correction } is deemed sufficiently small 
that the problem has converged 
to an optimal fit to the data.
Iterative Newton's method processes are the most common methods for solving the EIT problem, \matt{including } 
the popular open source Matlab package EIDORS 
\cite{eidorsuses}. This package allows the use of mesh based model-spaces and the selection of arbitrary data-spaces to simulate the \claire{linear and nonlinear forward EIT problem } and \claire{solve the } inverse EIT problem \claire{with a choice of various inverse solvers, such as the regularized inverse Newton's method approach described in Section \ref{sec:reginv}}. 


\section{Sensitivity Analysis Method and Protocol}

\matt{ The novelty of the proposed sensitivity analysis will now be explained with respect to 
the above mentioned components of model-space, data-space, and inversion method.  The model-space will be deliberately chosen to be small, identifying only the features or the resolution required, thereby accelerating the inverse problem.  The }\matthew{data measurement protocol will be compliant to maximize noise tolerance by rejecting } \claire{low sensitivity } \matthew{measurements from the protocol.  
}
\matthew{And to draw attention to the importance of a compliant protocol, we 
distinguish two new steps in the sensitivity analysis, namely the contact allocation and the identification of the measurement-space.  Contact allocation guarantees that the number of data measurements $D_0$ is {\em less than} the number of possible independent measurements $D_\mathrm{I}$, as per the criterion of a compliant protocol $D_I \gg D_0$.  The other new step is that of identifying the measurement-space, representing a deliberately large set of all possible $D_\mathrm{max}$ measurements, and defining the sensitivity vector for each measurement in order to set a figure-of-merit according to which measurements should be selected in the optimized protocol. 
}
\label{sec:sensitivity_analysis}

\subsection{Model-Space for Sensitivity Analysis}
\label{sec:model_space}
\color{black}

\matt{We first address the selection of model-space and model reference under the sensitivity analysis, with the intent of leveraging prior knowledge to }
\matthew{minimize the model-space dimension $M_0$. } \matt{
Following Section \ref{sec:intro_model_space}, the } first step is to determine a reference model from prior knowledge which will reduce the number of computational steps to converge on a final solution. 
This reference \matt{represents } 
the background \matt{conductivity } 
of the problem \matt{from which the final solution should be perturbatively related, if at all possible. }
An inhomogeneous \matt{reference } 
has been encouraged by Grychtol and Adler \matt{whenever prior knowledge of the system allows \cite{inhomogref}. } 
\matt{The next step of the sensitivity analysis is to identify a minimal number $M_0$ of model basis functions to describe the features that must be discerned so that the corresponding computational load can be minimized. }
\matt{Again following Section \ref{sec:intro_model_space}, }\matthew{ the minimal basis can either be determined according to a resolution criterion or by targeting specific features of interest, such as through proper orthogonal decomposition. \cite{basis} }

\begin{figure} [b]
    \centering
     \includegraphics[width=0.5\textwidth]{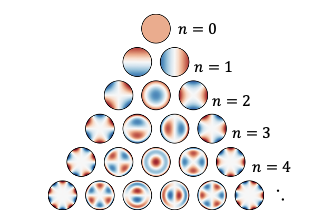}
    \caption{An orthonormal set of Zernike polynomial basis functions defined in Eq.\,\ref{eqn:zernikes} ranked  $n=0,1,2\dots N$ \matthew{ from low to high resolution
    } of polynomial order $n$ is chosen to reduce the model dimensionality over the circular area from \matthew{thousands } 
    of mesh points down to a handful $M\matthew{_0}$ of polynomials \matthew{
    that achieve the intended resolution of the final map. }  The number of independent voltage measurements $D\matthew{_0}$  is chosen, in turn, to match 
this number of basis functions
\matthew{ $D_0=M_0$, thereby creating an exactly determined problem and a square Jacobian matrix for inversion of the linear forward problem.} }
    \label{fig:zernfig}
\end{figure}

\matthew{The model-space that will be used for illustration of these concepts will be a circular domain with orthonormal polynomials as basis functions. To simplify the choice of reference } 
\matthew{in the present example, } we \matt{employ } 
a homogeneous constant background  \matt{reference } 
for \matt{compatibility with generic 2D systems.  } 
Given the circular symmetry of our test problem,  Zernike polynomials defined over a mesh will serve as conductivity basis functions for the model-space. 
The orthonormal Zernike polynomial functions are defined on a unit disk in terms of polar coordinates 
$Z_n^k(r,\theta)$, \matthew{and } represent increasing resolutions determined by a radial polynomial order $n$ and an azimuthal order $k$ \cite{zernikedefinition} as \matthew{illustrated } in Fig.\,\ref{fig:zernfig}. 
\begin{equation}
  Z_n^k(r,\theta) =
  \begin{cases}
     F_n^k R_n^{|k|}r \cos(k\theta) & \textrm{ for } k \geq 0, \\
     -F_n^k R_n^{|k|}r \sin(m\theta) & \textrm{ for } k<0,
  \end{cases}
  \label{eqn:zernikes}
\end{equation}
\matt{where } $R_n^{|k|}$ is defined as
\begin{equation}
    R_n^{|k|}(r) = \sum _{l=0}^{\frac{n-k}{2}} \frac{(-1)^l (n-l)! }{l! \left(\frac{n+k}{2}-l\right)! \left(\frac{n-k}{2}-l\right)!}r^{n-2 l}.
    \label{eqn:Rnm}
\end{equation}
The normalization prefactor $F_n^k$ 
normalizes the integral of the polynomials over the unit circle, 
\begin{equation}
    F_n^k = \sqrt{\frac{2(n+1)}{1+\delta_{k0}}}
    \label{eqn:Fnm}
\end{equation}
and contains the Kronecker delta function $\delta_{n0}.$ 
The Zernike polynomials satisfy the orthonormal condition, 

\begin{align*}
    \int_{\theta=0}^{2\pi}\int_{r=0}^1Z_n^k Z_{n'}^{k'} dr d\theta = \delta_{n,n'}\delta_{k,k'}.
\end{align*}

Using this polynomial basis, we can represent any conductivity map as \matthew{a} linear \matthew{combination } of Zernike polynomials:  
\begin{align}
    \sigma(r,\theta)=\sum_{n = 0}^{N}\sum_{k=-n,-n+2,...}^n m_{n,k}Z_n^k(r,\theta),
\end{align}
where $N$ is the highest polynomial order of the Zernike polynomials used. 
The weight of each polynomial is given by the coefficient \matt{$m_{n,k}$, whose components together define the model-space vector representation of the conductivity. }
For a given polynomial order $N$, the dimension $M_{\mathrm{Z}(N)}$ of the model-space up to and including that polynomial order is:
\eq{M_{\mathrm{Z}(N)}=\sum_{j=0}^N (j+1)=\frac{(N+1)(N+2)}{2}.}
\matt{In practice, the order $N$ of the polynomial will be chosen such that $M_{\mathrm Z(N)} \ge M_0$, } \matthew{where the $M_0$  polynomials for the model-space basis are chosen to suit the prior knowledge of the symmetry or needed resolution for the problem at hand. } 

\matthew{\subsection{Contact Allocation for Sensitivity Analysis}}
\label{sec:contact_allocation}

\matthew{A design aspect unique to the sensitivity analysis is contact allocation -- determining how many electrical contacts $C$ are necessary to provide the desired resolution }\claire{or features. }\matthew{For standard protocols the logic is normally reversed:  the number of contacts $C$ is given {\it a priori}, necessitating  $D_\mathrm{I}$ or more independent measurements according to \cref{data_independent}, where
the number of measurements $D_0 = r D_\mathrm{I}$ is either efficient ($r = 1$) or redundant ($r > 1$), as described in \cref{sec:intro_data_space}. For example, the reduced Sheffield protocol is an efficient protocol with $r = 1$ and the standard Sheffield protocol is redundant with redundancy factor $r=2$. In the sensitivity analysis proposed here, }
%
%
\matt{
the number of {\em possible} independent measurements $D_\mathrm{I}$ should {\em greatly exceed} the number of measurements in the protocol $D_0 = M_0$, whereby 
$D_0\ll D_\mathrm{I}.$ 
In practice, this is achieved by simply increasing the number of contacts, } \mattfinal{per }\cmmt{
Eq.\,\eqref{eqn:Dmax}. }
\matthew{One can explicitly invert Eq.\,\eqref{eqn:Dmax} and deduce the following expression for the minimum number of contacts $C$ needed to make $D_0$ independent data measurements:
\begin{equation}
    C(D_0) = \mathrm{ceil} \left(\tfrac{3}{2}+\sqrt{\tfrac{9}{4}+2D_0}\right) 
    \label{eq:C}
\end{equation}
where the ceiling function $\mathrm{ceil}(x)$ returns the smallest integer greater than or equal to $x$. According to \cref{sec:data_space}, a compliant protocol is defined by the condition $D_0 < D_\mathrm{I}$, so }\co{a compliance factor $c$ can be defined as follows: $D_\mathrm{I} = c\, D_0$ }.\matthew{ And the number of contacts to achieve this }\co{compliance }\matthew{ condition for an exactly determined Jacobian is therefore  $C(c\,M_0)$, thereby determining the number of contacts directly from the number of model-space basis functions $M_0$ as well as the } \co{ compliance factor $c$.
A compliance factor of $c \ge 3$ }\matthew{is recommended, and the larger $c$ is, the likelier the protocol can be optimized using only large-signal measurements.  Note that unlike the Sheffield protocol, }\matthew{there are no } \cmmt{ redundant 
measurements, }\matthew{as the sensitivity analysis relies on enhanced sensitivity rather than measurement redundancy to increase tolerance to noise.
} 

\subsection{Measurement-Space for Sensitivity Analysis} 
\label{sec:measurement_space}

\matt{Measurement-space is introduced as an essential component of the sensitivity analysis method proposed here.  
\matt{Borrowing from the data-space discussion in Section \ref{sec:intro_data_space}, recall that the total number of possible data measurements $D_\mathrm{max}$ } greatly exceeds the number of linearly independent measurements $D_\mathrm{I}$. } \cmmt{In this section, we }\matthew{shall }\cmmt{
define the space of all $D_\mathrm{max}$ possible measurements as the {\em measurement-space}, and introduce }
\matt{a metric whereby different measurements within this space can be quantified for their ability to distinguish model features, namely the {\em sensitivity vector}. } 

%
%

\matthew{The sensitivity analysis proceeds as follows. }
\matthew{Recalling the definition of a Jacobian, } we define sensitivity coefficients $J_{\co{a,\beta}}$ 
\matthew{that measure 
the sensitivity of each }\co{$a$ } \matthew{$ \in \{1,\dots D_\mathrm{max} \}$ data measurement to each } \co{$\beta$ }\matthew{$\in \{1,\dots M_0\}$ model element: } 
\eq{J_{\co{a,\beta}}& =\left.\frac{\partial F_{\co{a}} }{\partial m_{\co{\beta}}}\right. .}
\matthew{This definition is identical } 
to the Jacobian \matthew{matrix elements of Eq.\,\eqref{eq:matrix_element}, } except now the index \co{$a$ } spans {\em all} of the $D_\mathrm{max}$ possible data measurements, \matthew{not merely the $D_0$ data measurements that are used for the tomographic inversion.  We will refer to this larger matrix as the {\em sensitivity matrix} of dimension $D_\mathrm{max} \times M_0$, from which $D_0$ select rows will be drawn to form the $D_0 \times M_0$ Jacobian matrix for the linearized forward problem. } 
\matthew{These rows, then, form an important component of the analysis, as } each possible four-point measurement can be represented \matthew{as a vector }\co{ $\bm{\mu}_a$ }\matthew{in model-space expressing } 
the 
sensitivity of all model elements with respect to 
\matthew{the }\co{$a^\mathrm{th}$ }\matthew{data measurement: } 
\co{
\begin{equation}
   \bm{\mu}_a = \begin{bmatrix}
J_{a,1} & J_{a,2} & \dots & J_{a,M_0}
\end{bmatrix}.
  \label{eqn:sensvectors}
\end{equation}}
\matthew{The protocol optimization problem is now reduced to selecting the $D_0$ rows from this sensitivity matrix that represent the data measurements best suited to resolving the $M_0$ desired model features. } 


\matthew{Note that the } sensitivity vectors and sensitivity matrix are not limited to the Zernike polynomial basis. Other bases \cmmt{may } be preferred \cmmt{when given } prior knowledge of the system.  For example, proper orthogonal decomposition of a prior dataset \cmmt{allows } orthogonal features to be ranked from highest to lowest relevance in the hierarchy.  \cite{basis} \cmmt{O}ther \textit{a priori} resolution \matthew{requirements } or \matthew{targeted feature }
bases may be adapted, as well. 

\subsection{Data-Space for Sensitivity Analysis}
\label{sec:data_space}

\matt{The $D_0$ data } \matthew{measurements }\matt{ 
that comprise the data-space are }
\matthew{selected  from } 
\matt{the $D_\mathrm{max}$ total measurements in measurement-space
 } \matt{as an optimized subset }
\matt{that can span the desired }\matthew{$M_0$-element } \matt{ model-space. }   
\matthew{To simplify the inverse problem and eliminate the need for regularization, an exactly determined square matrix will 
be chosen for the data-space Jacobian such that $D_0 = M_0$. } 
\co{As previously emphasized in Section \ref{sec:intro_data_space}, }\mattfinal{when optimizing a protocol }
\co{it is important to avoid both measurements with low sensitivity and pairs of measurements with high similarity. } 

\mattfinal{To address these protocol criteria, first a {\em pruning threshold} is proposed to discard the lowest sensitivity measurements from consideration in the protocol.n 
} The magnitude of a sensitivity vector \matthew{$|\boldsymbol{\mu}_\alpha|$ } \mattfinal{quantifies } 
\matthew{ the sensitivity of that measured data value }\cmmt{
to }\matthew{any }\cmmt{changes in the model, where }
\matthew{vectors with greater }\cmmt{ magnitude 
are more noise-robust 
and lead to smaller posterior variance in the estimated model parameters from the inverse problem. }
\matthew{
Thus a ranked list of  
the $D_\textrm{max}$ sensitivity vectors can be pruned, whereby 
those shorter than a certain threshold }
 \claire{are considered low sensitivity and discarded. }
\mattfinal{Whereas pruning is not necessary for the moderately low number of contacts $C = 27$ considered here with $D_\mathrm{max}=52,650$ different measurements per \cref{eqn:Dmax}, it is anticipated to be a necessary step in geometries with significantly more  contacts. } 

\mattfinal{Second, a criterion is proposed whereby similar measurements can be avoided within a protocol. } 
\mattfinal{In model-space, this 
similarity is manifest as nearly parallel orientation of the respective sensitivity vectors. }
The more orthogonal the sensitivity vectors are, the more they represent independent measurements, giving a smaller posterior covariance of the model parameters in the inverse problem.
\claire{Optimal protocols are } \matthew{therefore } \cmmt{formed from } \matthew{sets of } 
\cmmt{ long, } \matthew{mutually } \cmmt{ orthogonal sensitivity vectors, } 
\mattfinal{with the figure-of-merit for 
such a protocol 
being the $D_0$-dimensional {\em sensitivity parallelotope volume} subtended by the sensitivity vectors of the constituent measurements. }
\mattfinal{For example, starting with }\co{two measurements, 
a parallelogram }\matthew{in model-space can be formed, }\mattfinal{representing the area $V_2$ }\matthew{subtended by the two 
}\mattfinal{respective }\matthew{sensitivity vectors } \mattfinal{$\boldsymbol{\mu}_1$ and $\boldsymbol{\mu}_2$. }
\mattfinal{This area will be maximized provided the two chosen vectors are as long and orthogonal as possible. }\co{An additional third measurement }\mattfinal{with sensitivity vector $\boldsymbol \mu _3$ will be most }\co{
sensitive to }\mattfinal{independent model-space components 
} 
\mattfinal{provided its } 
\co{perpendicular component with respect to the }\mattfinal{plane of the original }\co{
parallelogram }\mattfinal{is maximized, }\co{ yielding a three-dimensional volume $V_3$, {\it etc.} }
\cmmt{The product of the perpendicular components for all $D_0$ sensitivity vectors }\co{in a measurement protocol } \matthew{gives } \co{the following }\co{definition for the }\mattfinal{$D_0$-dimensional }\cmmt{sensitivity parallelotope volume: } 
\begin{equation}
    V_{D_0}=\prod_\alpha^{D_0} |\bm{\mu}_{\alpha\perp}|.
\label{eqn:compVolume}
\end{equation}
\matthew{The perpendicular component $\boldsymbol{\mu}_{\alpha\perp}$ of each successive measurement }
\matthew{can be determined from 
the } 
parallel component $\bm{\mu}_{\alpha\,||}$ of the $\alpha^\mathrm{th}$ 
sensitivity vector  relative to the subspace spanned by the \mattfinal{orthogonal components of the preceding } 
$\alpha - 1$ measurements:
\begin{equation}
\bm \mu_{\alpha\,||} = \sum_{i=1}^{\matthew{\alpha-1}
}\frac{\bm \mu_{i\perp} \cdot \bm \mu_\alpha}{|\bm \mu_{i\perp}|^2}\bm \mu_{i\perp}.
\end{equation}
The perpendicular component $\bm{\mu}_{\alpha\perp}$ \cmmt{is then:}
\begin{equation}\bm \mu_{\alpha\perp}=\bm \mu_\alpha-\bm \mu_{\alpha\,||}.
\end{equation}
\begin{figure}
    \centering
    \includegraphics[width=0.8\textwidth]{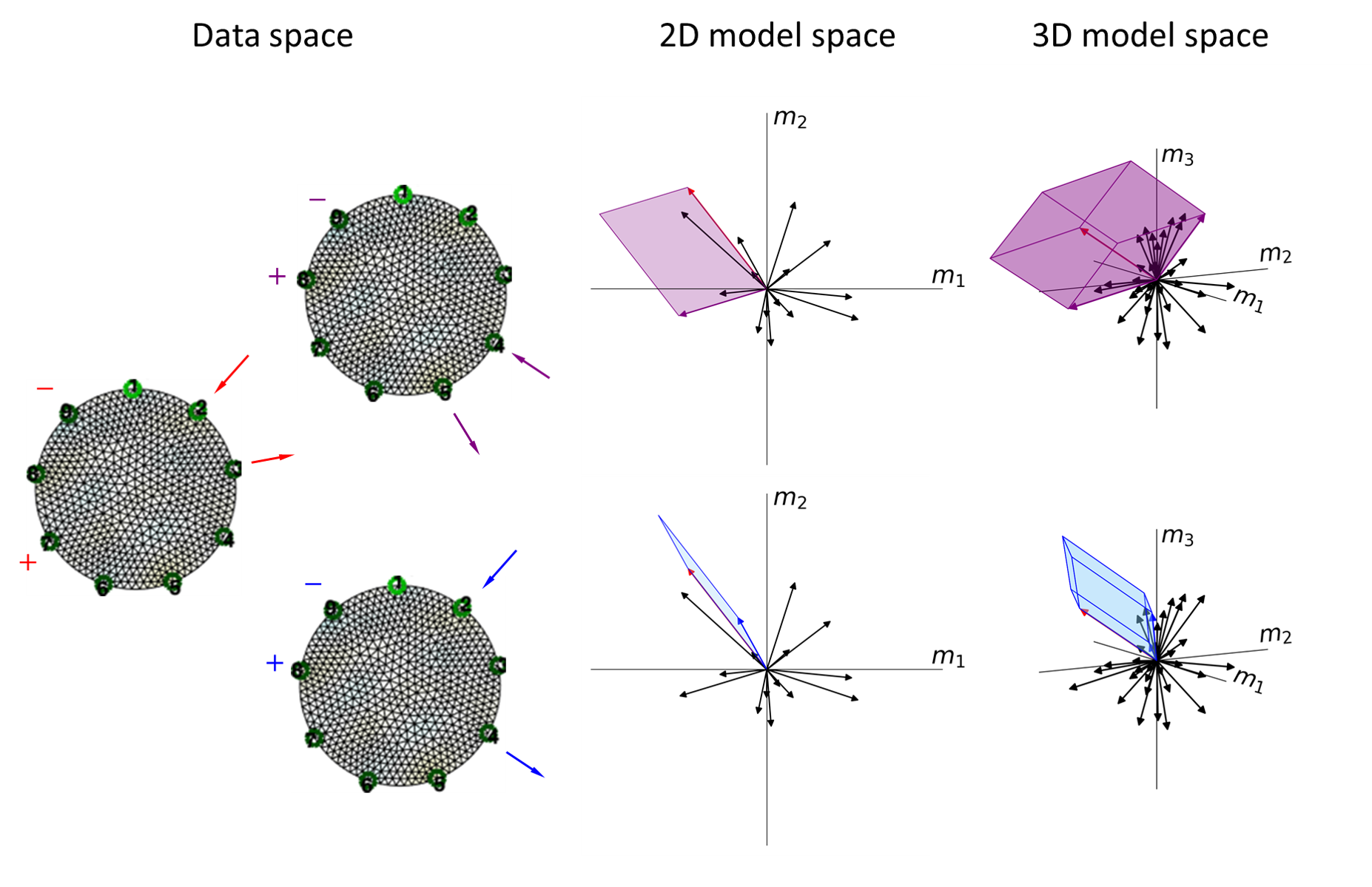}
    \caption{\matt{Sensitivity vectors and the sensitivity parallelotope volumes. To the left, measurement schematics show current (arrows) and voltage (+/-) terminals associated with red, purple, and blue measurement configurations (clockwise from left.) } With the first measurement selected (red), there are two choices for the subsequent measurement (purple, top; or blue, bottom). In the center column is the 2D model-space, where the axes are the model-space coefficients, $m_1$ and $m_2$, \matt{and the arrows represent sensitivity vectors corresponding to various possible measurements projected onto the $m_1$-$m_2$ plane. The parallelograns indicate the 2D sensitivity volume associated with a given pair of measurements, and } the top row is the better choice \matt{given the larger subtended area. }
    The rightmost column \matt{adds a third sensitivity vector to the set } showing the 3D sensitivity parallelotope volume. Compared to the botton row, the purple parallelotope volume again is a better choice of measurements because it maximizes \matt{both the magnitude and the orthogonality of the  sensitivity vectors in its set, thereby maximizing the parallelotope volume. }}
    \label{fig:parallelotope}
\end{figure}
\matthew{The sensitivity parallelotope volume 
in model-space
can be more formally calculated from the Jacobian made up of the sensitivity vectors } 
\matthew{of the $D_0$ selected measurements. The sensitivity parallelotope  volume associated with } 
the \matthew{Jacobian } matrix is: 
\begin{equation}
V_{D_0}=\sqrt{
\det{
(
\bm{JJ}
^\mathsf{T}
)
}
},
\label{eqn:Volume}
\end{equation}
%
\co{and is equivalent to that derived in \cref{eqn:compVolume}. }
The above sensitivity parallelotope volume figure-of-merit allows \matthew{an objective and generalized optimization of the tomographic protocol. }

\matthew{An example of the concept of parallelotope volume is illustrated in \cref{fig:parallelotope}. An initial measurement in data-space (red measurement, left) generates a corresponding sensitivity vector in 2D model-space (red vector, center).  There are two other candidate measurements shown (purple measurement, top left; blue measurement, bottom, left) which subtend a large-area parallelogram (top, center) or small-area parallelogram (bottom, center) in 2D model-space.  The measurement with larger 2D parallelogram area creates an overall larger sensitivity for the two-measurement protocol.  Analogously, in 3D model-space (right) the three vectors form two candidate protocols with large 3D parallelotope volume (top, right) or small 3D parallelotope volume (bottom, right.) Once again, the larger volume corresponds to an overall superior sensitivity for the protocol. }

\mattfinal{Several different } search algorithms \matthew{have been developed }\cmmt{ that }\matthew{maximize }
\cmmt{the sensitivity parallelotope volume. }
One such algorithm is a guided greedy search that works well for \matthew{smaller } contact numbers $C<50$. This algorithm builds up a protocol one measurement at a time, \cmmt{such that each } \matthew{newly added } \cmmt{measurement 
produces the largest possible volume }\matthew{with respect to all prior selected measurements in the protocol. } 
After \cmmt{a new } measurement is selected, a secondary check of \cmmt{all the other measurements ensures } that \cmmt{an alternate measurement will not } increase the sensitivity parallelotope volume \cmmt{further. When a more optimal } measurement is discovered during this double check, the checking process starts over again. Using this algorithm, and \co{a compliance factor of $c = 3$ }\mattfinal{corresponding to }\matthew{
a three-fold } 
increase in contact number \cmmt{over standard protocols, } we have \cmmt{generated measurement } protocols \cmmt{whose } \matthew{geometric averaged sensitivity vectors are } 
2 to 3 orders of magnitude \cmmt{larger than } \matthew{those of } the \cmmt{reduced } Sheffield protocol. \mattfinal{This leads to a commensurate increase in the noise tolerance such that 2 to 3 orders of magnitude more noise can be tolerated while yielding a faithful tomographic map of the conductivity. }
\matthew{This guided greedy search can be adapted to include }\cmmt{a } simulated annealing or Markov Chain Monte Carlo approach \matthew{to } add \mattfinal{randomness to the search } 
to \cmmt{improve } \matthew{search efficiency in the ever increasing measurement-space. } 
\matthew{Notably, } \cmmt{this 
} optimization process need 
only be computed once for each \cmmt{EIT problem 
prior to any inverse solving, } \mattfinal{enabling rapid real-time refresh of the tomographic inverse problem. } 

\subsection{Inverse Solution for Sensitivity Analysis}
\label{sec:inverse_sol}
To complete \cmmt{this } EIT sensitivity analysis method, \cmmt{one must solve the inverse problem}. As mentioned in \cref{sec:review}, this is often done via a Newton's iterative method, whereby the nonlinear EIT problem is solved 
\matt{linearly } with each iteration. In the following subsections, we will describe two approaches to a single iteration of a Newton's method solve. \cmmt{Either method can } be iterated to solve the nonlinear problem. However, we have found that if perturbations from \cmmt{the } initial solve guess are sufficiently small, \cmmt{a single iteration of the below methods provide sufficient convergence to solve the EIT inverse problem, producing equivalent results in a fraction $(1/100^\textrm{th})$ of the time taken by an iterative solve. }\mattfinal{For example, certain applications which might require rapid refresh, real-time tomography may benefit more from instantaneous inversion than having a few more significant digits of precision in the reconstruction. }
\subsubsection{Singular Value Decomposition (SVD):}
\label{sec:svd}
One method \cmmt{commonly used for }Newton's method iterations is (Truncated) Singular Value Decomposition (T)SVD solving. Singular value decomposition refers to the decomposition of the Jacobian into 3 unitary matrices of left and right singular vectors ($\bm{U},\bm{V}$) and a diagonal matrix of singular values ($\bm S$).
\eq{\bm{J}&=\bm{US V}^\mathsf{T}.
\label{eqn:SVD}}
Here, the number of non-zero singular values in $\bm S$ is \cmmt{equivalent to the rank of } the Jacobian. For each \cmmt{singular value}, $\bm U$ has a column to represent the contributing linear combination of data-space elements and a column of $\bm V$ stores the corresponding linear combination of model-space elements. 
The columns of $\bm{U}$ are the eigenvectors of $\bm{JJ}^\mathsf{T}$, while the columns of $\bm{V}$ are the eigenvectors of $\bm{J}^\mathsf{T}\bm{J}$
In (T)SVD solving, \cmmt{the decomposition in  \cref{eqn:SVD} forms }the generalized inverse of $\bm J$: 
\eq{
\bm{J}^{-g}=\bm{V}\bm{S}^{-1}\bm{U}^\mathsf{T}.
\label{eqn:Jg}}
\cmmt{The combination of Eqs. \eqref{eqn:inverse} and \eqref{eqn:Jg} produce the following expression for the inverse solution }\matthew{in the presence of noise: }
\eq{
\bm{m}&=\bm{V}\bm{S}^{-1}\bm{U}^\mathsf{T}{(\bm d+\bm n)},}
where $({\bm d+\bm n}=\bm d^{\textrm{obs}})$ is the observed data that includes noise \cmmt{and $\bm{m}$ is the best possible reconstruction from the noisy data.} 

\cmmt{In the case that some }singular values in $\bm S$ are metrically or numerically equivalent to zero, \cmmt{truncation is used to }replace the small singular values with zeros. \cmmt{If the small singular values were kept for inverse solving, }their reciprocals \cmmt{within $\bm{S}^{-1}$ } \matthew{would } approach infinity and cause widely inaccurate inverse solutions due to noise. \cmmt{Truncation remedies this by ensuring that noisy information is ignored}. \cmmt{To indicate truncation to $k$ singular values, the singular value matrix is written as $S_k$ for some $k<\textrm{rank}(\bm{S})$}:
\eq{\bm{S}_k=\begin{bmatrix}
s_1 &  &  & & &  \\
 & s_2 &  & & &  \\
 &  & \ddots & & &  \\
 &  &  & s_k& &  \\
 & & & & 0 & \\
 &  & &  & & \ddots
\end{bmatrix}.}
The \cmmt{truncation point } $k$ is selected \cmmt{with } the discrete Picard condition \cite{TSVD} to maintain a solution that is not consumed by noise. This condition creates Picard coefficients from the observed data and the $\bm{U}$ matrix from singular value decomposition \cite{Kantartzis}: 
\eq{\bm{p}=\bm U^\mathsf{T}(\bm d+\bm n).}
\cmmt{The Picard coefficients show a decay that decreases upon reaching the }noise level $\bm{n}$. The discrete Picard condition states that the decay \cmmt{of these } Picard coefficients must be faster than the decay observed in the singular values $s$ found in the diagonals of $S_k$. \cmmt{If there exists an index $k$ such that }the singular value $s_k$ is decaying faster than the Picard coefficient $p_k$, truncation becomes necessary. This \matthew{(T)}SVD \matthew{solution } 
is a fast and efficient way to form the inverse. In the nonlinear regime, the problem is iterated with either the same or varying truncation points. In this work, we have written Matlab code to perform non-iterative \matthew{(T)}SVD solves for our arbitrary \cmmt{polynomial }basis function model-spaces. 

\cmmt{Not all EIT problems can be solved within the aforementioned linear regime, providing a necessity for iterative solving. } In practice, the same measurement protocol is used for each iteration of a nonlinear solve. However, \cmmt{with a } sufficiently large library of \cmmt{optimized }sensitivity measurement protocols, \cmmt{one } can start with the best protocol for the reference guess and \cmmt{switch } to an improved protocol based \cmmt{on the results of an initial linear solve. }Then, the new protocol would be used in the iterations. 
\subsubsection{Regularized Inverse}
\label{sec:reginv}
As mentioned in \cref{sec:review}, the EIT community has developed multiple methods for improving iterative Newton's method solves of the EIT problem, \cmmt{ one of which is by use of a regularized inverse. This popular approach, as described by Graham {\it et al.} \cite{hyperparam} uses }the regularized inverse to solve the inverse problem as:
\eq{
\bm{ m}&=(\bm{J}^\mathsf{T}\bm{ W J}+\lambda\bm{G})^{-1}\bm{J^{\mathsf{T}}} \bm W{ (\bm d+\bm n)},}
where $\bm{G}$ is a regularization matrix determined by the type of regularization used, \cmmt{such as flattening or } smoothing, and $\bm{W}$ is a matrix that models noise. The hyperparameter $\lambda$ controls \cmmt{the extent to which } regularization is applied to the problem. 

\mattfinal{Either the regularized inverse approach or the (T)SVD approach previously described }
\cmmt{
can be used with optimized sensitivity measurement protocols and reduced dimensionality model-spaces. }For the sake of \matthew{simplifying } the examples in the following section, we chose to use our \mattfinal{(T)}SVD based solver within the linear, single iteration regime. 


\section{Results and Discussion}
\label{sec:results}
In this section the signal to noise advantage of our sensitivity measurement protocols \cmmt{is illustrated through the application of } EIT to \matthew{phantom models for problems in } biomedical imaging and structural engineering. In \cmmt{the } first example, a medical imaging scenario \cmmt{shows } how the selection of data-space measurements via \cmmt{the } sensitivity \cmmt{analysis method }
results in a \matt{factor of $30\times$ higher noise tolerance compared to } 
the reduced Sheffield protocol. In 
\matt{the language of linear algebra, } \cmmt{the sensitivity optimized measurement protocols 
} lead to a posterior model covariance matri\matthew{x } that ha\matthew{s } both low variance and low covariance values. 
In the second example, the same analysis \cmmt{is shown } for a structural engineering scenario in which prior knowledge is utilized in the selection of basis functions \matt{leading to an even greater factor of $1,000\times$ increase in noise tolerance compared to } \cmmt{reduced } \matt{Sheffield. } The EIDORS EIT package was used 
to solve the forward problem and plot the reconstructions seen in  \cmmt{the following examples}. \cmmt{The inverse solves were performed using our own linearized SVD solver as described in section \ref{sec:svd}. }

\subsection{Biomedical Example}
One popular \mattfinal{proposed } \matthew{biomedical }  application for EIT technology \cmmt{
is chest imaging. } For example, 
\matthew{measurements of stroke volume and ejection fraction \mattfinal{for a heart }
can diagnose cardiac }\claire{conditions such as heart failure and coronary syndrome \cite{heart1}\cite{heart2}\cite{chesteit}\cite{RaoHeart}. }
To simulate this, \cmmt{a circular sample with a homogeneous conductivity serves }\matthew{to model }\cmmt{a chest cavity cross section, and a smaller }\matthew{included  circle 
with }\co{20}\matthew{\% higher } \cmmt{conductivity represents the blood-filled heart. For the sake of this demonstration of methods, other chest cavity features were ignored. In practice, features of varying conductivity, such as the lungs, would be included as an inhomogeneous background reference. }

\cmmt{The steps of our sensitivity analysis were applied to this scenario.  First, the homogeneous circular reference and a desired 
}\matthew{spatial } \cmmt{resolution } \matthew{corresponding to }\cmmt{ $M_0=27$ }\matthew{polynomials } \cmmt{were used to form a Zernike 
basis model-space. For } \matthew{the exactly determined inverse problem, } \cmmt{an equivalent 
data-space dimensionality }\matthew{of }\cmmt{
$D_0=27$ independent measurements }\matthew{is required.  From \cref{eq:C}, }\cmmt{this would require at least }
\matthew{$C(D_0) = 9$ } \cmmt{contacts, } \matthew{chosen to be } \cmmt{equally spaced. However, as recommended in } \matthew{Section \ref{sec:contact_allocation}, }\co{a compliance factor of $c=3$ }\matthew{ is used to determine a better contact number for maximizing sensitivity, namely $C(c\,D_0)=27.$ } 
Next, a standard 
EIT forward solver \cite{eidors} \matthew{is used }
to calculate the sensitivity coefficients \matthew{that make up } 
the \cmmt{sensitivity vectors for } \matthew{each of the $D_\mathrm{max} = 52,650$ possible } measurements. 
 \cmmt{Then, } our sensitivity optimization algorithm generated an optimal data-space of 
\matthew{$D_0$ } \mattfinal{of these } measurements \cmmt{to maximize the sensitivity parallelotope }\matthew{volume. }
\mattfinal{In order to compare with }
\matthew{standard EIT methods, } 
\cmmt{a $D_\mathrm{I}=27$ measurement data-space was also formed from the }\matthew{ $C(D_\mathrm{I})
=9$ contact } \cmmt{reduced Sheffield protocol. }\matthew{Using the parallelotope volume as a basis for comparison, the sensitivity protocol had a }\mattfinal{27-dimensional } \matthew{parallelotope volume that was $10^{28}$ times larger than that of the Sheffield protocol, corresponding to an average enhancement of the sensitivity by a factor of $11\times$ per measurement.}

\matthew{To show the consequences of this enhanced sensitivity in the tomographic inversion, }
this EIT example 
\cmmt{was solved } with our own linear, 
\mattfinal{(T)SVD } solver 
\matthew{per } \cref{sec:inverse_sol}. To avoid biased inverse solving, simulated data \cmmt{was generated on a very }fine mesh and the \cmmt{Zernike } polynomial model-space \cmmt{was used } for \matthew{inversion. 
Gaussian }\cmmt{noise of varying }\matthew{amplitude with }\cmmt{ standard deviation $\eta$ was added to the simulated data }\matthew{to emulate realistic experimental conditions. }
After solving noisy data sets for both our protocol and the reduced Sheffield protocol, \cmmt{results were compared to assess the ability of } each protocol to differentiate between a ``normal heart" and \cmmt{an } ``enlarged heart". 

As a metric 
\mattfinal{for differentiating these two cases, } 
a 
distinguishability \matthew{criterion } $z$ was \cmmt{determined}, \matthew{inspired by } 
the work of Adler \cite{adjacent}. 
\matthew{The distinguishability } metric \cmmt{treats the average result }of many \mattfinal{noisy } EIT inverse solves 
as a vector in model-space \matthew{ $\overline{\bm{m}}
$} \co{: 
\begin{equation}
\overline{\bm{m}}=E[\bm{m}],
\end{equation}}
\mattfinal{where $E[\cdot]$ is the ensemble expected value. }
\matthew{The variance of all the noisy inversions }\mattfinal{of a given case }\matthew{ 
form an ellipsoidal surface on model-space described by the covariance matrix }\co{$\bm{\Sigma}$:
\eq{\bm{\Sigma}=\begin{bmatrix}
\sigma^2_1 & \sigma^2_{1,2} & \dots & \sigma^2_{1,M_0} \\
\sigma^2_{2,1} & \sigma^2_2 & \dots  &  \sigma^2_{2,M_0}\\
\sigma^2_{3,1} & \sigma^2_{3,2} & \ddots  &  \sigma^2_{2,M_0} \\
\vdots & \vdots & \vdots  &  \sigma^2_{M_0} \\
\end{bmatrix}.}
\mattfinal{whereby } the variances and covariances 
are, respectively:
\begin{equation}
\sigma^2_{\beta}=E\left[||\bm{m}_\beta-\overline{\bm{m}_\beta}||^2\right]
\label{eqn:variance}
\end{equation}
\eq{
\sigma^2_{\beta\gamma}=
E\left[(\bm{m}_\beta-\overline{\bm{m}}_\beta)(\bm{m}_\gamma-\overline{\bm{m}}_\gamma)\right],
\label{eqn:cov}}
} 
for $\beta, \gamma \in \{ 1,\cdots M_0\}$. 
\mattfinal{When comparing two distinct ensembles of noisy inversions, in this example a ``normal heart" with average model-space vector $\overline{\bm{m}}_{\textrm{n}}$ } and an ``enlarged heart" with average \matthew{$\overline{\bm{m}}_{\textrm{l
}}$}\cmmt{, } one can compute the \co{component of the } standard deviations $\Delta m_\mathrm{n}$ and $\Delta m_\mathrm{l}$ \matthew{projected }\cmmt{along } the direction of the difference between the two \matthew{$\overline{\bm{m}}_\mathrm{l}-\overline{\bm{m}}_\mathrm{n}$ }\co{as follows: } 
\co{\eq{\Delta m_\mathrm{n,l}^2=\bm{\hat{u}}^T \bm{\Sigma}_\mathrm{n,l} \bm{\hat{u}},}
where }\mattfinal{the unit vector $\bm{\hat{u}}$ pointing from $\overline{\bm{m}}_\mathrm{n}$ to $\overline{\bm{m}}_\mathrm{l}$ is 
\eq{\bm{\hat{u}}=\frac{\overline{\bm{m}}_\mathrm{l}-\overline{\bm{m}}_\mathrm{n}}{||\overline{\bm{m}}_\mathrm{l}-\overline{\bm{m}}_\mathrm{n}||}.}}
%
\cmmt{Distinguishability } $z$ \cmmt{can then be defined as }\mattfinal{the distance in model space between the two cases relative to the combined standard deviations of the two cases, } 
\begin{equation}
    z = \frac
    {||\overline{\bm{m}}_{\textrm{l}}-\overline{\bm{m}}_{\textrm{n}}||}
    {\sqrt{
    \Delta m_\mathrm{n}^2
    +
    \Delta m_\mathrm{l}^2.}}
    \label{eqn:dist}
\end{equation}
For $z \gg 1$ the distinguishability is very high, \cmmt{and the resulting reconstructed images }\mattfinal{of the two cases }\cmmt{are clearly different } \mattfinal{ from each other }\matthew{in spite of the noise. 
} \mattfinal{Furthermore, the reconstructed images for each case is indistinguishable by eye from the noiseless reconstruction. } However, when the value of $z$ is less than 1, \matthew{noise has overwhelmed the discernable difference between the two cases, and } the patterns 
are 
\matthew{no longer reliably } 
distiguishable, \mattfinal{nor do the reconstructed images exactly resemble the noiseless reconstruction. } 

\begin{figure}[ht]
    \centering
    \includegraphics[width=1.0\linewidth]{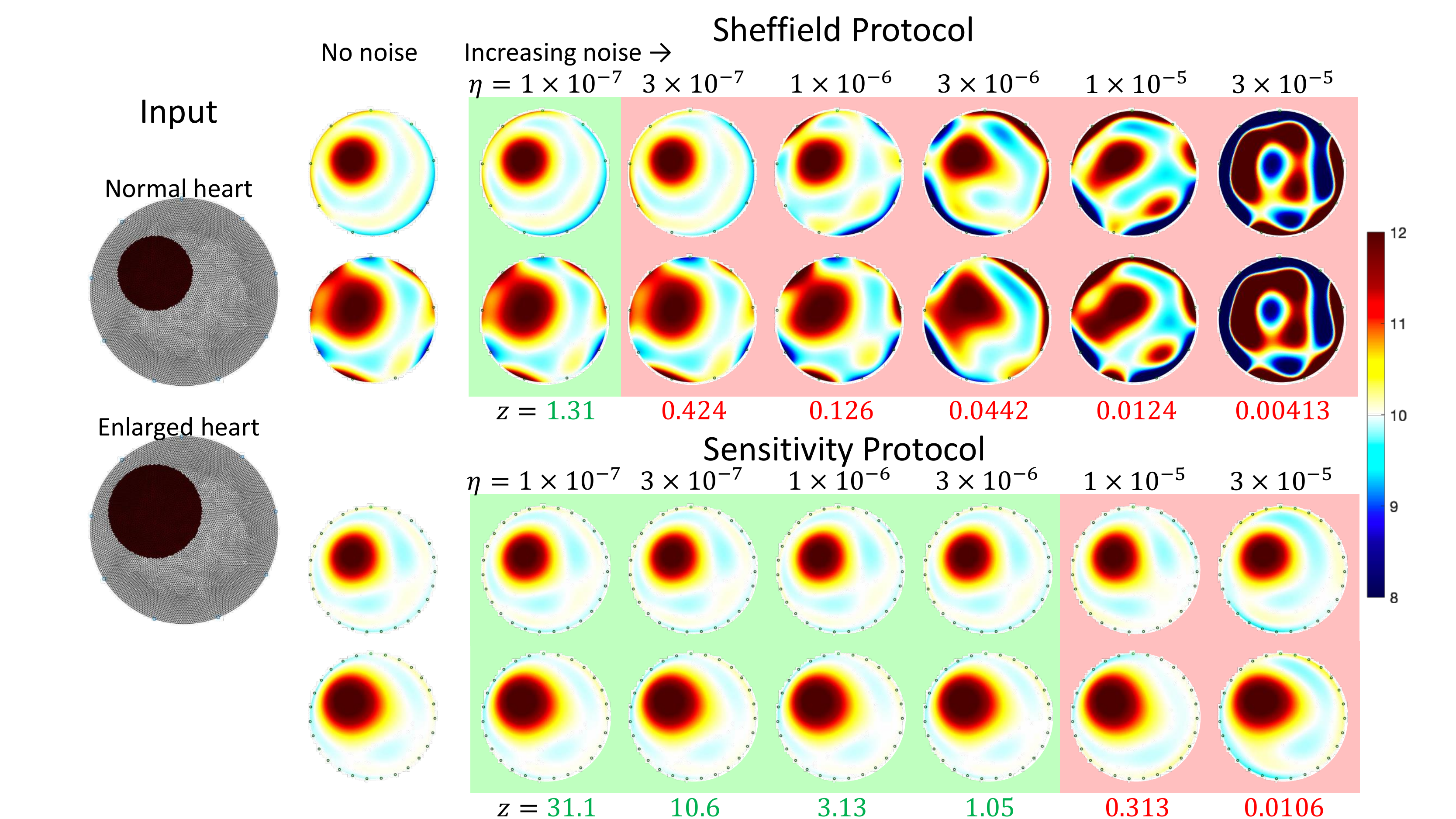}
    \caption{Comparing reconstruction quality and distinguishability $z$ between a ``normal heart'' and ``enlarged heart'' in a simplified 2D model of a chest cavity. 
    A factor of $30\times$ larger noise is tolerated by the sensitivity protocol compared to the reduced Sheffield protocol. 
    Input resistivity maps and mesh are on left (black and white) and reconstructed images with\cmmt{out added } noise \mattfinal{immediately to their right. }
     Increasing noise level $\eta$ from left to right shows that the \cmmt{reduced }Sheffield protocol (top-right rows) faithfully reproduces the noiseless reconstruction only up to \mattfinal{a noise level of } $\eta = \co{1\times}10^{-\co{7}}$
     \mattfinal{(green background represents inversions that {\em can} be statistically distinguished; red for indistinguishable or distorted inversions). } The sensitivity protocol, on the other hand (bottom-right rows), tolerates a factor of $30\times$ higher noise level $\eta=\co{3}\times 10^{-\co{6}}$.
    The distinguishablility criterion $z$ below each pair of reconstructions highlights reliable reconstructions $z > 1$ (green) from reconstructions whose difference in features between ``normal'' and ``enlarged'' is less than the noise variation (red)., 
    Conductivity color scale on far right.}
    \label{fig:heart}
\end{figure}

In Fig.~\ref{fig:heart} 
\matthew{the } sensitivity protocol \matthew{is observed to show }
a 
noise \matthew{tolerance } \cmmt{advantage } of approximately \matthew{a factor of $30\times$ larger noise compared to the standard protocol: } 
\cmmt{from the images and $z$ values, it is clear that the sensitivity protocol can }\matthew{accurately invert with a noise level of } 
up to $\eta=3\times 10^{-\co{6}}$, \cmmt{whereas } the Sheffield protocol can only handle noise 
up to $\eta=1\times 10^{-\co{7}}$. 
\cmmt{For the } inverse solves \cmmt{used in this example}, the singular values were not truncated during solving. Higher levels of noise can be endured at the loss of information when using truncation. 
\matthew{But even under truncation, } the observed signal to noise improvement of \matthew{the sensitivity protocol } \mattfinal{relative to the Sheffield protocol }
remains. 

\subsection{Structural Engineering Example}

\begin{figure}[ht]
    \centering
    \includegraphics[width=1.0\linewidth]{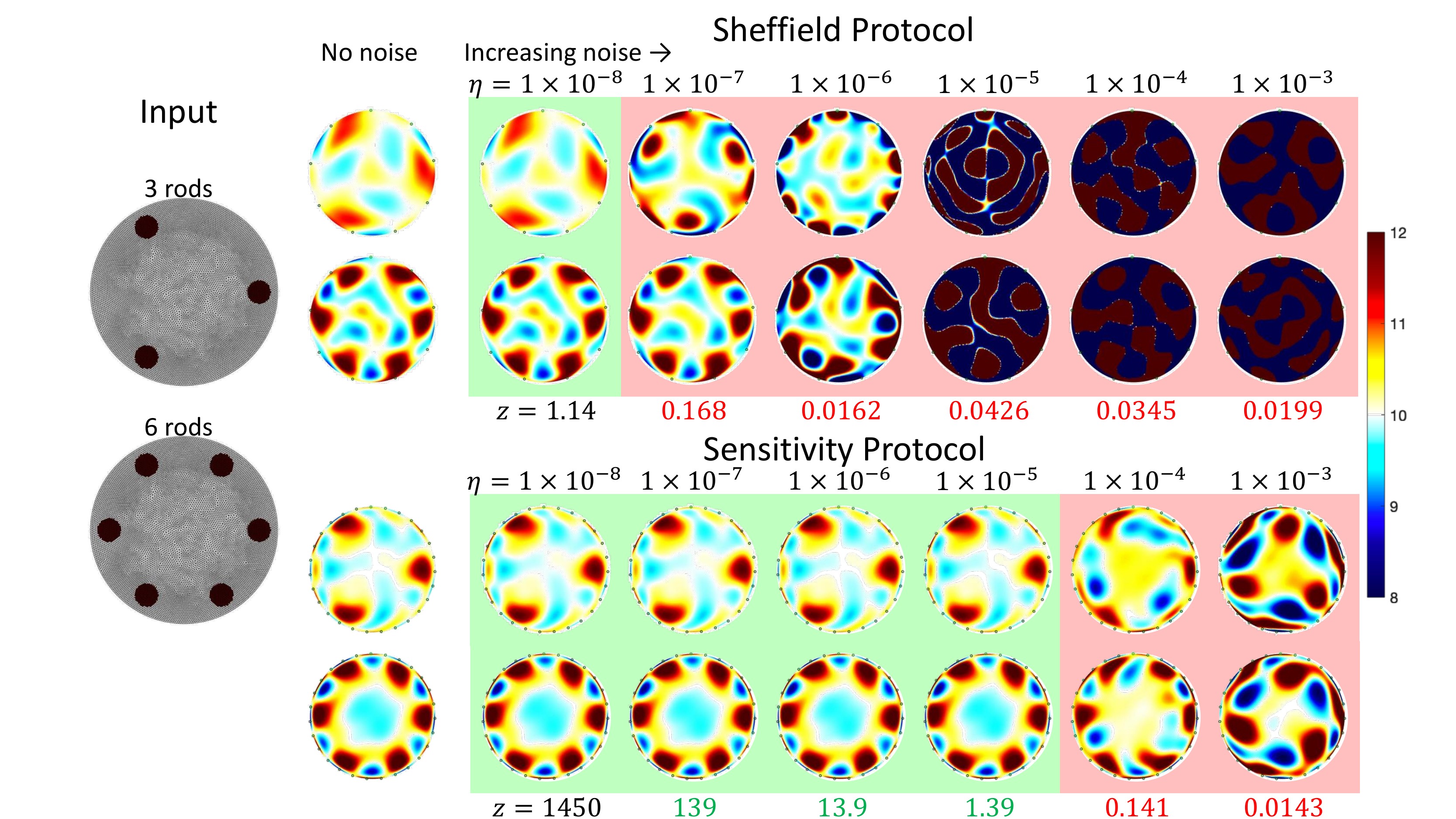}
    \caption{Comparing reconstruction quality and distinguishability $z$ between 3 and 6 reinforcement rods in the cross-sectional model of a concrete pillar. A factor of $1000\times$ larger noise is tolerated by the sensitivity protocol compared to the traditional Sheffield protocol.  Input resistivity maps and mesh are on left (black and white) and reconstructed images with no noise \mattfinal{immediately to the right. }  
    Increasing noise level $\eta$ from left to right shows that the \cmmt{reduced } Sheffield protocol (top-right rows) faithfully reproduces the noiseless reconstruction only up to $\eta = \co{1\times} 10^{-\co{8}}$ \mattfinal{(green background represents inversions that {\em can} be statistically distinguished; red for indistinguishable or distorted inversions).} The sensitivity protocol, on the other hand (bottom-right rows), tolerates a factor of $1,000\times$ higher noise level $\eta=1\times 10^{-5}$. The distinguishablility criterion $z$ below each pair of reconstructions highlights reliable reconstructions $z > 1$ (green) from reconstructions whose difference in features between ``3 rod" and ``6 rod" is less than the noise variation (red). 
    Conductivity color scale on far right.}
    \label{fig:concrete}
\end{figure}

\matthew{The } second demonstration of our \cmmt{sensitivity analysis method }
\matthew{will use } a concrete pillar \matthew{as a structural engineering example. } 
In practice, \cmmt{cylindrical concrete pillars are reinforced with equally spaced }conducting \matthew{wrought iron } reinforcement rods. \matthew{Maintenance inspections of historical structures or enforcement of building codes may require identifying the amount of reinforcement in a finished structure. } 
To simulate this \matthew{problem, the cross-section of a concrete pillar is modeled as }
a homogeneous circular sample with either 3 or 6 ``rods'' \matthew{at regular intervals within its volume. }
\cmmt{The model-space }\matthew{for } \cmmt{this problem } was 
strategically \mattfinal{constructed using }
\matthew{
the }
\cmmt{prior knowledge that the } \matthew{solution would have 3-fold rotational symmetry, and selected $M_0 = 27$ polynomial basis functions under that constraint. }
\matthew{Of } the standard Zernike polynomials, the first $M_0 =27$ polynomial \cmmt{functions } 
\matthew{that satisfy the 3-fold symmetry requirement 
require sampling polynomials up to order $n = 12.$ \mattfinal{(For example, \cref{fig:zernfig} depicts up to order $n = 5$, which contains only the first seven of the 3-fold symmetric polynomials.) }
\matthew{Following } 
model-space 
\matthew{selection, the contact allocation again requires $C(u\,M_0)=27$ equally spaced contacts}. The next step requires calculation of all $D_\mathrm{max}=52,650$ sensitivity vectors  
for all possible measurements. }
\cmmt{For the } \matthew{final } \cmmt{ 
step, the data-space was selected } \matthew{from this measurement-space by optimizing the parallelotope volume. }
Data-space selection 
used the same algorithm as in the previous example to produce a 
\claire{parallelotope }volume with a \matthew{remarkable $10^{64}$ increase relative to the reduced Sheffield protocol, which breaks down to an average noise tolerance a factor over $200\times$ greater 
per measurement for the same number of measurements. }

Linear (T)SVD 
\matthew{is again used to solve the inverse problem. }
Due to the magnitude of the singular values \cmmt{from the Jacobian matrices for } this particular model-space, all inverse solves, both of \matthew{the }
sensitivity protocol and the reduced Sheffield protocol, were truncated to 18 singular values. 
\cmmt{Then, the } \matthew{distinguishability  
$z$ for various noise levels was } 
calculated to compare the ability to \matthew{differentiate } 3 and 6 rods under both 
protocols 
\matthew{The Gaussian noise with } standard deviation $\eta$ was \matthew{once again } simulated for 1000 \cmmt{different trials}. 
The results 
in Fig.~\ref{fig:concrete} 
show that 
\matthew{the } sensitivity protocol can tolerate 
noise of 3 orders of magnitude \matthew{larger }
than the \cmmt{reduced } Sheffield protocol: 
the reduced Sheffield protocol patterns no longer \cmmt{determine } the correct reconstruction and are not distinguishable in the presence of noise $\eta\ge 1 \times 10^{-7}$. 
In contrast, the sensitivity protocol is \cmmt{acceptable } until noise of $\eta=1\times 10^{-4}$.

\section{Conclusions}
This manuscript \matthew{details } 
a 
sensitivity analysis 
\matthew{for optimizing } 
the EIT inverse-problem \cmmt{ for noise robust measurements and improved computational speed. The above analysis demonstrates five 
steps for the careful selection of model-space and data-space elements, }\matthew{leveraging } 
prior information and \mattfinal{introducing } the sensitivity parallelotope volume figure of merit. The sensitivity analysis 
\matthew{procedure } chooses a reduced dimensionality model-space in which to solve the EIT problem, determines how many contacts can yield the necessary model resolution \claire{or features of interest}, defines  
a measurement-space \mattfinal{which contains } the sensitivity vectors \matthew{that describe all possible measurements, } selects the optimal data-space measurements as a subset of this measurement space, and finally, solves the inverse problem. 

\mattfinal{With the proposed sensitivity analysis method, the goal if rapid-refresh EIT inversion is brought within reach because of two operating in tandem. } 
\mattfinal{First, the choice of a polynomial basis greatly reduces } the dimensionality of the model-space which, in turn, greatly reduces the computation time of the problem \mattfinal{in comparison to the standard regularization problem with a high density of mesh points}. 
\mattfinal{Second, the introduction of the
sensitivity parallelotope volume permits the optimization of} 
\cmmt{measurement protocols for maximum sensitivity, resulting in }\matthew{significantly } \cmmt{improved }robustness to noise in comparison to conventional protocols. \cmmt{In the examples presented to illustrate our methods, we observed a factor of up to $1,000\times$ improvement in noise tolerance compared to standard EIT protocols. }

\section*{References}
\bibliography{ref.bib}
\bibliographystyle{unsrt}
\end{document}